\documentclass[oupdraft]{bio}
\usepackage{hyperref}

\usepackage{comment}
\usepackage{enumerate}
\usepackage{caption}
\usepackage{subcaption}
\usepackage{rotating}
\usepackage[normalem]{ulem}
\usepackage{algorithm}
\usepackage{verbatim}
\usepackage{rotating}
\usepackage{algpseudocode}

\usepackage{xr}
\renewcommand{\v}[1]{\mathbf{#1}}
\newcommand{\vs}[1]{\boldsymbol #1}
\newcommand{\minimize}[1]{\underset{#1}{\text{minimize}}}
\newcommand{\argmin}[1]{\underset{#1}{\text{argmin}}}



\begin{document}

\title{Simultaneous detection and estimation of trait associations with genomic phenotypes}

\author{Jean Morrison$^\ast$ $\dag$, Noah Simon$\dag$, Daniela Witten$\ddag$ \\[4pt]
\textit{$\dag$ Department of Biostatistics\\ 
$\ddag$ Departments of Statistics and Biostatistics \\ 
University of Washington, Seattle WA, 98195, USA} \\[2pt]
{jeanm5@uw.edu}}

\markboth%
{J Morrison, N Simon, D Witten}
{Detection and estimation  of associations with genomic phenotypes}

\maketitle
\footnotetext{To whom correspondence should be addressed.}

\section*{Abstract}
 Genomic phenotypes, such as DNA methylation and chromatin accessibility, 
 can be used to characterize the transcriptional and regulatory activity of DNA within a cell. 
 Recent technological advances have made it possible to measure such phenotypes very densely.  
 This density often results in spatial structure, in the sense that measurements at nearby sites are very similar.

 In this paper, we consider the task of comparing genomic phenotypes across experimental conditions, cell types, or  disease subgroups. 
  We propose a new method, Joint
Adaptive Differential Estimation (JADE), which leverages the spatial structure inherent to genomic phenotypes. JADE simultaneously estimates
smooth underlying group average genomic phenotype profiles, and detects 
regions in which the average profile differs between groups. 
 We evaluate JADE's performance in several biologically plausible simulation settings. We also consider an application to the detection of regions with differential methylation between mature skeletal muscle cells, myotubes and myoblasts.

\section{Introduction}
\label{sec:intro}

During the past decade, it has become possible  to measure \textit{genomic phenotypes}, or local properties of DNA such as DNA methylation, chromatin accessibility, copy number variation, and histone modification. 
 These genomic phenotypes can be measured very densely, and in some cases even at  single-nucleotide resolution. For example, DNA methylation proportion can be measured at nearly every CpG using bisulfite sequencing, and chromatin accessibility can be measured at every nucleotide with DNase-seq. Neighboring nucleotides may belong to the same functional unit. Thus,  genomic phenotypes often have similar values at nearby genomic positions.

Unlike the DNA sequence itself, genomic phenotypes may change dynamically as a function of cellular activity, developmental stage, and environment. 
 This motivates us to compare genomic phenotypes across experimental conditions or clinically-defined categories. 
  This task is made challenging by the fact that, for most genomic phenotypes, 
  we cannot pre-specify all relevant functional units 
 and test each unit for an overall difference across groups.
 For instance, for epigenetic features, many  potentially relevant regulatory regions are not annotated. Furthermore, in exploratory studies, it is not always known which classes of genetic elements, such as promoters or enhancers, should be considered. 

When pre-specified functional units are not available, 
 a simple option  for comparing genomic phenotypes across conditions is to  perform a separate test
at each locus.
The majority of existing methods  take such an approach, for instance using  logistic regression, Fisher's exact test, or t-tests \citep{Akalin2012,Stockwell2014,Jaffe2012,Hebestreit2013}, sometimes followed by 
a multiplicity correction that considers spatial structure.  
These procedures identify differential regions by merging contiguous sites with large test statistics.
 Such tests tend to be under-powered, as they fail to borrow strength across neighboring loci.  

In contrast, 
the BSmooth method of \cite{Hansen2012} and the WaveQTL  method of \cite{Shim2013} are two-step procedures that leverage the spatial structure of the genomic phenotypes.
 BSmooth first smooths the data, and then uses the smoothed data to calculate a $t$-statistic at each site. 
 Differential regions are then identified by merging contiguous sites with large $t$-statistics.
 WaveQTL requires the genome to be divided into pre-specified bins. A  hierarchical Bayesian regression is performed in order to generate a bin-level test statistic, as well as estimates of association between the data and the outcome at different spatial scales.

In this paper, we propose \emph{joint adaptive differential
estimation} (JADE), a one-step approach for differential estimation and testing of genomic phenotypes. 
JADE  is a penalized likelihood-based approach which
simultaneously estimates smooth average-group profiles and identifies
regions of difference between groups. 
By combining these two tasks into a single step, JADE can adaptively share information both across loci and between
groups, leading to improved power to detect differential regions without the need for pre-specified functional units of interest.
 When the grouping variable has more than two levels, JADE finds regions where at least one group differs from the rest, and within those differential regions performs local clustering of profiles.

 The rest of this paper is organized as follows. 
In Section~\ref{sec:problem},  we introduce the underlying model, and formulate JADE as the solution to a convex optimization problem.
In Section~\ref{sec:solution}, we introduce a custom algorithm that can be used to efficiently solve the JADE optimization problem. 
In Section~\ref{sec:sims}, we explore the performance of JADE, relative to existing methods, in a simulation study.
In Section~\ref{sec:methylation} we apply JADE to publicly available methylation data from the ENCODE project. The Discussion is in  Section~\ref{sec:disc}.

\section{Problem Formulation}
\label{sec:problem}

Consider a categorical trait, $X \in\left\{1,\ldots,M\right\}$, such as disease status or tissue type, coded numerically for convenience. 
 We wish to associate this trait with a genomic phenotype, $\v{Y}=(Y_1,\ldots,Y_p)^T$, measured at positions $s_1<s_2<\ldots<s_p$ along the genome. 
 
For a given value of $X$, we assume that $\v{Y}$ varies smoothly as a function of genomic position, 
\[
f_{m}\left(s_j\right) = E\left[Y_{j}\middle|X=m\right].
\]
Here the function $f_m$ represents the mean genomic phenotype profile for the $m$th class.
   If all $m$ profiles are identical at site $s_j$ ($f_m\left(s_j\right) = f_{m'}(s_j)$ for all $m\neq m^{\prime}$), then there is no association between the mean of $Y_j$, the genomic phenotype at the $j$th position, and the categorical trait $X$. 
 If $f_m\left(s_j\right) \neq f_{m'}(s_j)$ for some $1 \leq m < m' \leq M$, then there is an association between the mean of $Y_j$ and $X$.  {Our goal is to identify \emph{differential regions}, or contiguous blocks of associated sites.}
   A very similar framework was considered in \cite{Shim2013}.

In what follows, we assume that we have $n$ independent 
observations  of $(X,\v{Y})$, denoted $(x_1,\v{y}_1),\ldots,(x_n,\v{y}_n)$. 
We now introduce some notation that will be used throughout this paper. 
  Let $N_m$ denote the number of observations with $x_i =m$, so that $N_1 + \ldots + N_M = n$. Let $\bar{y}_{mj} \equiv \sum_{i: x_i = m}  y_{ij} / N_m$, and let $\bar{\v{y}}_{m} \equiv\left(\bar{y}_{m 1}, \ldots, \bar{y}_{m p}\right)^{\top}$.
 Furthermore, we let $\theta_{mj}\equiv f_m\left(s_j\right)$, and $\vs{\theta}_m \equiv \left(\theta_{m1},\ldots,\theta_{mp}\right)^{\top}$. In what follows, unless otherwise specified,  the letter $i$ will index the $n$ observations, $m$ will index the $M$  values of the categorical trait $X$, and $j$ will index the $p$ genomic positions of $Y$. 

\subsection{Example}
\label{sec:example}
\begin{figure}[!p]
\centering
	\begin{subfigure}[b]{0.3\textwidth}
		\includegraphics[width=\textwidth]{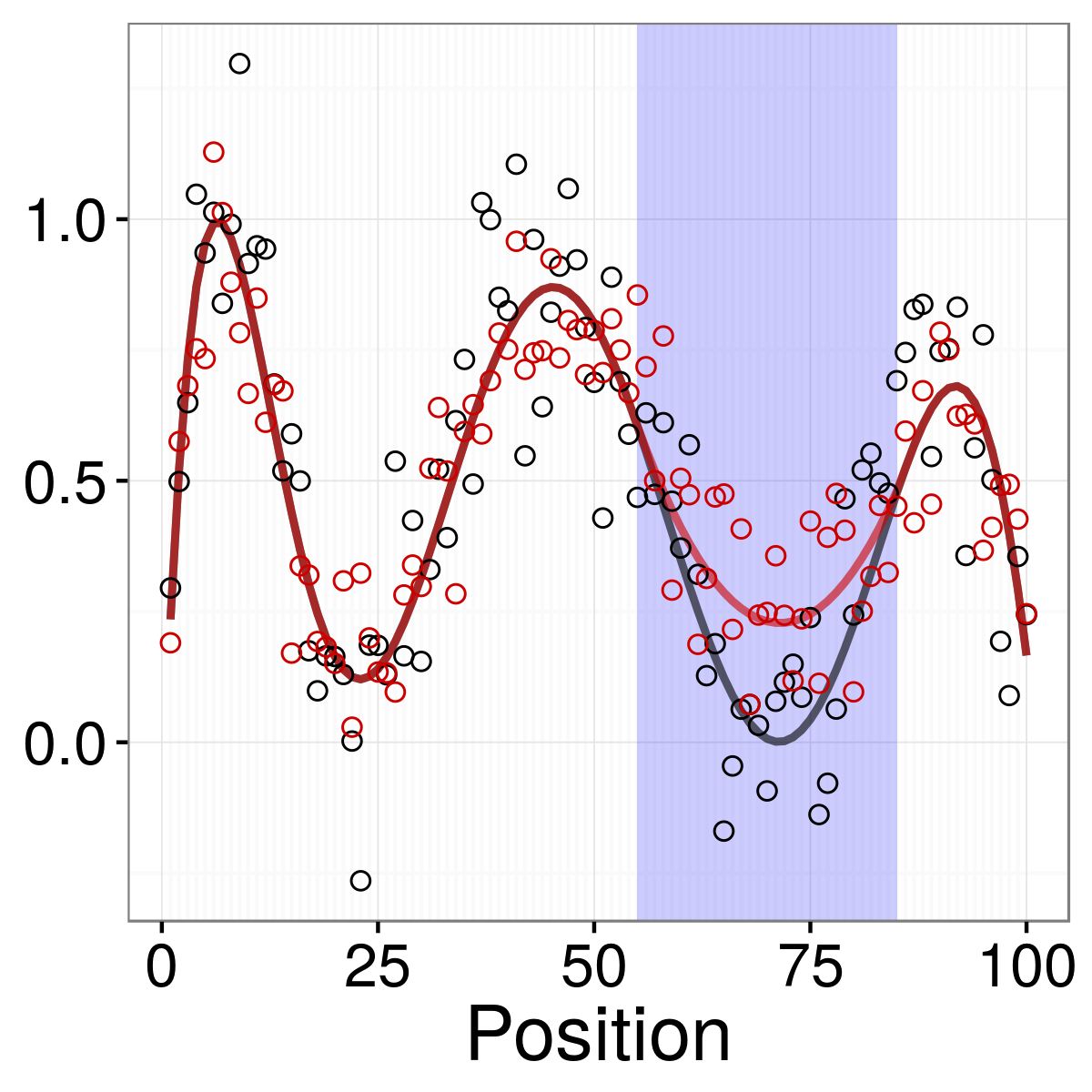}
		\caption{True Profiles}
		\label{panela}
	\end{subfigure}
	\begin{subfigure}[b]{0.3\textwidth}
		\includegraphics[width=\textwidth]{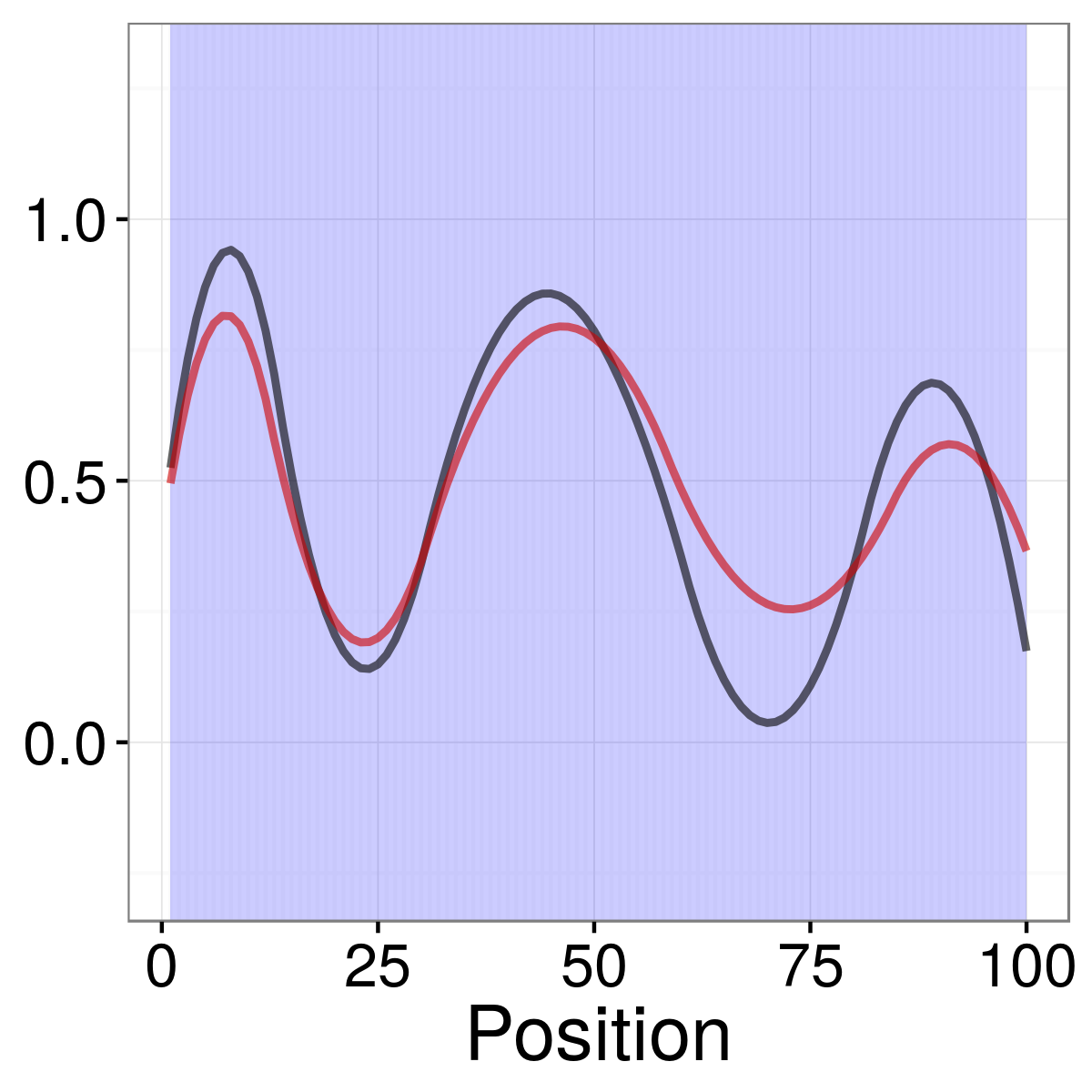}
		\caption{Smoothed Profiles}
		\label{panelb}
	\end{subfigure}
	\begin{subfigure}[b]{0.3\textwidth}
		\includegraphics[width=\textwidth]{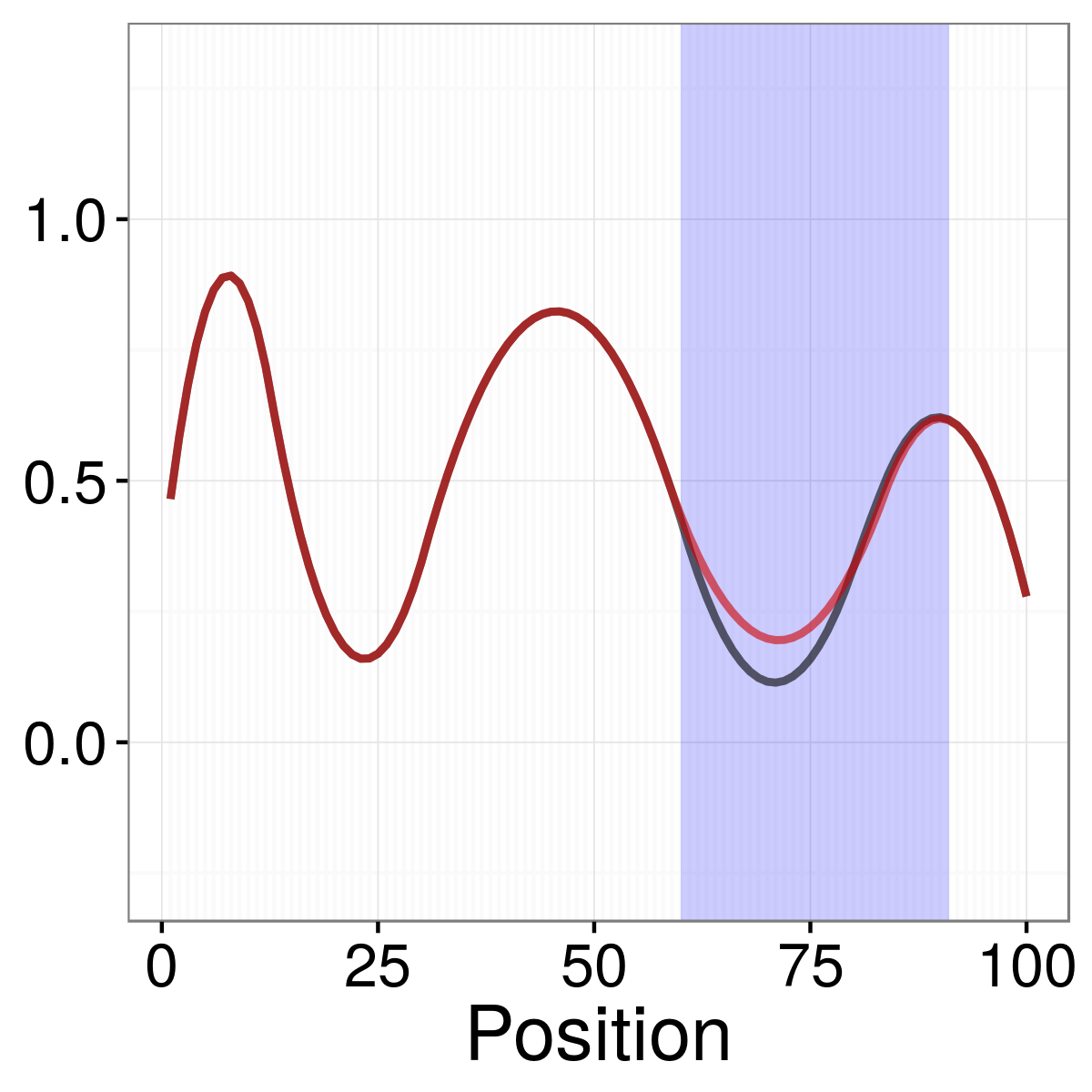}
		\caption{JADE Fit}
		\label{panelc}
	\end{subfigure}
\caption{An illustration of the toy example described in Section~\ref{sec:example}. In Figure~\ref{panela}, red and black data points are generated as normal observations with mean given by the corresponding colored lines. Blue shading in \ref{panela} indicates the region in which the two true profiles are not identical. In Figure~\ref{panelb}, profile estimates are obtained by smoothing the two groups separately. These profiles are separated over the entire region.  In Figure~\ref{panelc}, profile estimates are obtained from JADE. The small region in which the estimated profiles differ is shaded in blue. The detected region largely overlaps the true region of difference.}
\label{toyfig}
\end{figure}

We {illustrate} JADE with a simple toy example. 
 In each of two groups, we simulate  a  quantitative genomic phenotype at a series of evenly spaced positions, $s_1,\ldots,s_p$. The  data are generated as an overall group-specific mean curve, plus independent normal errors, as shown in Figure~\ref{panela}.  The two group-specific mean curves differ only for $s_j \in [55,85]$. 
 
 We first consider estimating the mean curves   by separately smoothing the data corresponding to each of the  two groups.  
As is shown in   Figure~\ref{panelb},  the two estimated profiles are somewhat different at nearly every location.

In contrast, the results from applying JADE  to this data are shown  in Figure~\ref{panelc}. JADE simultaneously smooths the data in each group, and penalizes the differences between the two estimated mean curves. Therefore, JADE can approximately recover the differential region shown in Figure~\ref{panela}.

Of course,
the data that we encounter in real biological problems, such as the application studied in Section~\ref{sec:methylation}, are more complicated than the toy example shown in Figure~\ref{panela}. Real data are often characterized by unevenly spaced positions $s_1,\ldots,s_p$; sites for which a subset of groups are missing measurements; and non-constant variance of the genomic phenotype measurements. As we describe in the following sections,  JADE is able to accommodate all of these characteristics. 

\subsection{Penalties to Induce Structure}
JADE combines two tasks: (i) estimation of a smooth mean curve within each group; and (ii)  fusion of the mean curves across groups. 
 Here we use the term \emph{fusion} to describe JADE's ability to provide  mean curve estimates that are identical across multiple groups at a particular genomic position. That is, if our estimates of $f_m(s_j)$ and $f_{m'}(s_{j})$ are identical for some $m \neq m'$, then we say that the estimated mean curves for the $m$th and $m'$th classes are \emph{fused} at position $s_j$.

We briefly discuss the application of existing penalized regression methods to the two aforementioned  tasks.

\subsubsection{Smoothing a Genomic Phenotype}
\label{sec:smooth}
Consider the task of smoothing a single observation of a genomic phenotype, $\v{y}_i \in \mathbb{R}^p$, measured at {(potentially unevenly spaced)} positions $s_1 < \ldots < s_p$. 
 Given weights $a_1,\ldots,a_p$, we consider the optimization problem  
\begin{align}
 \minimize{f}\ \left\{ \frac{1}{2}\sum_{j=1}^{p} a_{j}\left(y_{ij} - f(s_{j})\right)^{2} + \lambda P\left(f\right)\right\}.
\label{eq:smooth}
\end{align} 
The smoothed estimate, $\hat{f}$, minimizes the sum of two terms: a goodness-of-fit term between $y_{ij}$ and $f(s_j)$, and a penalty term that discourages a rough or complex $f$. The penalty parameter $\lambda$ controls the relative importance of these two terms.
There are a number of options for $P(\cdot)$, such as a smoothing spline penalty \citep{Reinsch1971} or an  $\ell_1$ trend filtering penalty \citep{Kim2009,Tibshirani2014}. 

Trend filtering induces piecewise polynomial estimates for $\hat{f}$, of pre-specified order $k$, with adaptively chosen knots. The choice of $k$ is guided by the characteristics of the data at hand: for instance, trend filtering with $k=0$ \citep[also known as the fused lasso; see][]{Tibshirani2005} is appropriate for the piecewise constant structure of copy number data \citep{Tibshirani2008}; while $k=2$ is  appropriate for relatively smooth DNA methylation data. 
Trend filtering is \emph{locally adaptive}, in the sense that it can be used to fit a curve that is very smooth in one region of the domain and very rough in another; this is discussed extensively in  \cite{Tibshirani2014}. This property is very attractive within the context of analyzing messy, heterogeneous, and heteroskedastic biological data. Consequently, in what follows, we take $P(\cdot)$ in \eqref{eq:smooth} to be a trend filtering penalty.

For convenience, we now switch to using vector notation. 
 The $\ell_1$ trend filtering estimate, $\hat{\vs{\theta}}$, is the solution to the optimization problem 
\begin{align}
 \minimize{\vs{\theta} \in \mathbb{R}^p}\ \left\{  \frac{1}{2}\sum_{j=1}^{p} a_{j}\left(y_{ij} - \theta_{j}\right)^{2} + \lambda\left\|\v{D}^{k+1,s} \vs{\theta}\right\|_1 \right\}, \label{eq:tf}
\end{align}
 where $\v{D}^{k+1, s}$ is the $(p-k-1)\times p$ discrete $(k+1)$th derivative matrix, the entries of which depend on both $k$ and the spacing of $s_1,\ldots,s_p$. The specific form of this matrix is detailed further in Section~\ref{sec:TF} of the supplementary material [SM] available at \textit{Biostatistics} online, and in \cite{Tibshirani2014}. 

The weights $a_1,\ldots,a_p$ in (\ref{eq:smooth}) and \eqref{eq:tf} can account for heterogeneity in the variance of $y_{ij}$. Setting  $a_1=\ldots=a_p$ gives equal weight to each position. Setting $a_{j}$ proportional to the inverse of an estimate of the variance of $y_{ij}$ gives less weight to positions with lower quality data.

\subsubsection{Fusing Genomic Phenotypes}\label{sec:fuse}
Now consider the task of fusing $n$ observations of a genomic phenotype, $\v{y}_1,\ldots,\v{y}_n \in \mathbb{R}^p$ --- that is, we seek to encourage the estimated means to be identical at a given site.  The \emph{convex clustering} estimates, $\hat{\vs{\theta}}_1,\ldots,\hat{\vs{\theta}}_n$, solve the optimization problem \citep{Pelckmans,Hocking2011,Heinzl2014} 
\begin{align}
 \minimize{\vs{\theta}_1,\ldots,\vs{\theta}_n \in \mathbb{R}^p}  \left\{ \sum_{i=1}^{n} \frac{1}{2}\sum_{j=1}^{p} a_{ij}\left(y_{ij} - \theta_{ij}\right)^{2} + \gamma \sum_{i<i^{\prime}} \left\|\vs{\theta}_{i}-\vs{\theta}_{i^{\prime}}\right\|_{q} \right\}. \label{eq:clustering}
\end{align}
In \eqref{eq:clustering}, $a_{ij}$ is a weight for the $j$th locus in the $i$th observation.

For $\gamma$ sufficiently large, the $\ell_q$ penalty in \eqref{eq:clustering} will encourage similarity between $\hat{\vs{\theta}}_i$ and $\hat{\vs{\theta}}_{i^{\prime}}$. In particular, if $q=2$, then when $\gamma$ is large, the entire vectors 
$\hat{\vs{\theta}}_i$ and $\hat{\vs{\theta}}_{i^{\prime}}$ will tend to be identical, or completely fused at all sites. In this case, 
 the set of observations for which $\hat{\vs{\theta}}_i$ are identical can be interpreted as clusters.   
  In contrast, if $q=1$, then a large value of $\gamma$ will encourage individual elements $\hat\theta_{ij}$ and $\hat\theta_{i'j}$ to be identical. This amounts to  fusing  the vectors $\hat\theta_i$ and $\hat\theta_{i'}$ at a subset of the sites. 

\subsection{Joint Smoothing and Comparison with JADE}
\label{sec:JADE}

Recall from the beginning of Section~\ref{sec:problem} the problem set-up: each observation belongs to one of $M$  categories,  $N_m$ denotes the number of observations within the $m$th category, and $\bar{\v{y}}_m \in \mathbb{R}^p$ denotes the mean of the observations of the genomic phenotype within the $m$th category.

Our goal is to estimate a mean genomic phenotype profile, $\hat{\vs{\theta}}_1,\ldots,\hat{\vs{\theta}}_M \in \mathbb{R}^p$, for each of the $M$ categories. We want each mean profile to be smooth, and for the $M$ mean profiles to be identically equal to each other for many of the loci $s_1,\ldots,s_p$. To do this, we 
  combine the  smoothing and fusion penalties seen in \eqref{eq:tf} and \eqref{eq:clustering}
  into a single convex optimization problem.

  The JADE estimator is defined as the solution to the convex optimization problem
\begin{align}
\minimize{\vs{\theta}_1,\ldots,\vs{\theta}_M \in \mathbb{R}^p}  \left\{ \sum_{m=1}^{M} \frac{N_m}{2}\left\| \v{A}_{m}\left(\bar{\v{y}}_m - \vs{\theta}_m\right) \right\|_{2}^{2}  + \lambda\sum_{m=1}^{M}\left\| \v{D}^{k+1, s}\vs{\theta}_m\right\|_{1} + \gamma \sum_{m < m^{\prime}}\left\|\vs{\theta}_m - \vs{\theta}_{m^{\prime}} \right\|_{1} \right\}. \label{eq:jadeproblem_full}
\end{align}
This minimization consists of three terms: a weighted sum of squared residuals, a sum of $\ell_1$ trend filtering penalties, and a clustering penalty.  
 When the non-negative tuning  parameter $\lambda$ is sufficiently large, the trend filtering penalty encourages each mean profile to be smooth. 
 Equation~\eqref{eq:jadeproblem_full} could be modified to allow each of the $M$ groups to have its own smoothness tuning parameter, $\lambda_1, \dots, \lambda_M$. For simplicity we use a single common parameter.
 
  When the non-negative tuning parameter $\gamma$ is sufficiently large, the clustering penalty encourages many of the $p$ sites to have exactly the same value in the $m$th and $m'$th mean profiles, for $m \neq m'$.
 In fact, when $\gamma$ is large enough, some of the $p$ sites will have $\hat\theta_{1j}=\ldots=\hat\theta_{Mj}$; these can be interpreted as regions of the genome where the mean profile is constant across the $M$ groups. Thus, JADE simultaneously identifies regions of the genome in which the genomic phenotype is associated with the categorical variable $X$, and estimates smooth average profiles for each group. It accomplishes this in an efficient way that borrows strength across nearby sites, without performing a separate test at each site in the genome. 
Selection of $\lambda$ and $\gamma$ in \eqref{eq:jadeproblem_full} is discussed in Section~\ref{sec:cv}.

In \eqref{eq:jadeproblem_full}, 
$\v{A}_{m}$ are $p\times p$ diagonal weight matrices. These can be used to account for the fact that the elements of  $\bar{\v{y}}_m$ may have non-constant variance across the $p$ sites, perhaps due to varying numbers of reads across the genome.
 Furthermore, if no data are available for the $j$th site in the $m$th group, then the $j$th diagonal element of $\v{A}_m$ can be set to zero.

\section{Solving the JADE Optimization Problem}
\label{sec:solution}

\subsection{An Alternating Direction Method of Multipliers Algorithm for JADE}\label{sec:algo}

The JADE optimization problem  \eqref{eq:jadeproblem_full} is convex,  so in principle, it can be solved with general-purpose convex solvers, such as \verb=SDPT3= \citep{Tutuncu2003} or \verb=SeDuMi= \citep{Sturm1999}. However, these solvers do not scale well to genome-sized problems. Therefore, we have developed an efficient custom \emph{alternating direction method of multipliers}  \citep[ADMM;][]{Boyd2011} algorithm for solving \eqref{eq:jadeproblem_full}.  

 Our algorithm relies on the key observation by \cite{Tibshirani2014} that the trend filtering penalty matrix $\v{D}^{k+1, s}$ can be decomposed as $\v{D}^{k+1,s} = \v{D}^{1}\tilde{\v{D}}^{k, s}$, where $\v{D}^{1}$ is the $(p-k-1)\times (p-k)$ first difference operator, and $\tilde{\v{D}}^{k, s}$ is a $(p-k)\times p$ scaled $k$th-order difference operator. Details of these two matrices are provided in  Section~\ref{sec:TF} of the SM.

 Using this decomposition, we can re-write \eqref{eq:jadeproblem_full} as
\begin{align}\label{eq:constr}
\minimize{\vs{\theta}_{1},\ldots,\vs{\theta}_{M},\vs{\beta}_1,\ldots,\vs{\beta}_M,\vs{\alpha}_1, \ldots, \vs{\alpha}_M} &  \left\{ 
\sum_{m=1}^{M} \frac{N_m}{2}\left\| \v{A}_m \left(\bar{\v{y}}_m - \vs{\theta}_{m}\right) \right\|_{2}^{2}  + \lambda\sum_{m=1}^{M}\left\| \v{D}^{1}\vs{\alpha}_m\right\|_{1} + \gamma \sum_{m < m^{\prime}}\left\| \vs{\beta}_m - \vs{\beta}_{m^{\prime}} \right\|_{1} \right\}\\
\text{subject to} \qquad &\tilde{\v{D}}^{k, s}\vs{\theta}_m = \vs{\alpha}_m, \ \ \vs{\theta}_m=\vs{\beta}_m, \qquad m=1,\ldots,M. \nonumber
\end{align}
The scaled augmented Lagrangian for this problem is
\begin{align}
L(\vs{\theta},\vs{\alpha},\vs{\beta}, \v{u}) = & \sum_{m=1}^{M} \frac{N_{m}}{2}\left\| \v{A}_m(\bar{\v{y}}_m - \vs{\theta}_{m})\right\|_{2}^{2} + \lambda \sum_{m=1}^{M}\left\| \v{D}^{1}\vs{\alpha}_m \right\|_1 + \gamma\sum_{m<m^{\prime}}\left\| \vs\beta_m-\vs\beta_{m^{\prime}}\right\|_{1} \nonumber \\
 &+ \frac{1}{2}\sum_{m=1}^{M}\rho_{\alpha m}\left\| \tilde{\v{D}}^{k, s} \vs\theta_m - 
\vs\alpha_m + \v{u}^{(\alpha)}_m\right\|_{2}^{2}  
 + \frac{\rho_{\beta}}{2}\sum_{m=1}^{M}\left\| \vs\theta_m  - 
\vs\beta_m + \v{u}^{(\beta)}_m\right\|_{2}^{2} ,
\label{eq:scaug}
\end{align}
where  $\vs\theta \equiv (\vs\theta_1^\top,\ldots,\vs\theta_M^\top)^\top$, $\vs\alpha\equiv(\vs\alpha_1^\top,\ldots,\vs\alpha_M^\top)^\top$, and $\vs\beta\equiv(\vs\beta_1^\top,\ldots,\vs\beta_M^\top)^\top$. In \eqref{eq:scaug},  $\v{u} \equiv (\v{u}_1^\top,\ldots,\v{u}_M^\top)^\top$
 is a vector of dual variables, where $\v{u}_m \equiv \left( \left( \v{u}_m^{(\alpha)} \right)^\top, \left(\v{u}_m^{(\beta)} \right)^\top \right)^\top$ for $\v{u}_{m}^{(\alpha)} \in \mathbb{R}^{p-k}$ and $\v{u}_{m}^{(\beta)}\in \mathbb{R}^p$. 
  The dual variables are broken into multiple components in order to allow for different step sizes, $\rho_{\alpha1},\ldots,\rho_{\alpha M}$ and $\rho_\beta$, as this leads to faster convergence. In our implementation, we adjust the step sizes adaptively; details are in Section~\ref{sec:stepsize} of the SM.
 
\begin{algorithm}
\caption{ADMM Algorithm  For Solving  the JADE Optimization Problem (\ref{eq:jadeproblem_full}) \label{alg:simple}}
\begin{enumerate}
\item  Initialize $\vs\beta_{1}, \dots , \vs\beta_{M}$ as solutions to (\ref{eq:jadeproblem_full}) with $\gamma=0$.
\item For $m=1,\ldots,M$,  initialize $\v{u}_m = 0$  and $\vs\alpha_m =\tilde{\v{D}}^{k,s}\vs\beta_m$.
\item Iterate until the convergence criteria described in Section~\ref{sec:stepsize} of the SM are satisfied: 
\begin{enumerate}
\item For $m=1,\ldots,M$, update
 \begin{align*}
 \vs\theta_m \longleftarrow & \left(N_{m} \v{A}_{m}^{\top}\v{A}_m + \rho_{\alpha m} \left( \tilde{\v{D}}^{k, s}\right)^{\top}\tilde{\v{D}}^{k, s} + \rho_{\beta}\v{I}\right)^{-1} \\
 & \cdot \left(N_{m}\v{A}_{m}^{\top}\v{A}_{m}\bar{\v{y}}_m + \rho_{\alpha m}\left(\tilde{\v{D}}^{k, s}\right)^{\top}\left(\vs\alpha_m-\v{u}_{m}^{(\alpha)}\right) + \rho_{\beta}\left(\vs\beta_{m}-\v{u}_{m}^{(\beta)}\right)\right).
 \end{align*}
\item
For $m=1,\ldots,M$, update 
$$\vs\alpha_{m} \gets \argmin{\vs\alpha_m} \left\{ \frac{1}{2}\left\| \vs\alpha_{m} - \left(\tilde{\v{D}}^{k, s}\vs\theta_{m} + \v{u}_{m}^{(\alpha)}\right) \right\|_{2}^{2} + \frac{\lambda}{\rho_{\alpha m}}\left\| \v{D}^{1} \vs\alpha_{m}\right\|_{1}\right\}.$$ 
\item
  For $m=1,\ldots,M$, update
$$\vs\beta_m \gets \argmin{\vs\beta_m} \left\{ \sum_{m=1}^{M} \frac{1}{2}\left\| \vs\beta_{m} - \left(\vs\theta_{m}+ \v{u}_{m}^{(\beta)} \right)\right\|_{2}^{2} + \frac{\gamma}{\rho_{\beta}}\sum_{m<m^{\prime}}\left\| \vs\beta_{m}-\vs\beta_{m^{\prime}} \right\|_{1}\right\}.$$ 
\item
For $m=1,\ldots,M$, update the dual variables by setting $$\v{u}_{m}^{(\alpha)} \gets \v{u}_{m}^{(\alpha)} +  \tilde{\v{D}}^{k, s}\vs\theta_{m} -\vs\alpha_{m}, \hspace{10mm} \label{ualpha}
 \v{u}_{m}^{(\beta)} \gets \v{u}_{m}^{(\beta)} +   \vs\theta_m  - \vs\beta_{m}. \label{ubeta}$$ 
\item Update the step sizes $\rho_{\alpha 1}, \dots, \rho_{\alpha M}$ and $\rho_{\beta}$ as described in Section~\ref{sec:stepsize} of the SM, and rescale the dual variables by setting
\[
\v{u}_m^{(\alpha)} \gets \v{u}_m^{(\alpha)} \cdot \rho_{\alpha m}^{old}/\rho_{\alpha m} \qquad \v{u}_m^{(\beta)} \gets \v{u}_m^{(\beta)} \cdot \rho_{\beta}^{old}/\rho_{\beta}.
\]
\end{enumerate}
\end{enumerate}
\end{algorithm}

The ADMM algorithm corresponding to the scaled augmented Lagrangian \eqref{eq:scaug} is given in   Algorithm~\ref{alg:simple}. 
The initialization in Step 1 simply amounts to solving a separate $\ell_1$ trend filtering problem for each $\vs\beta_m$, $m=1,\ldots,M$.
 The update  in Step 3(b) involves solving a fused lasso problem; this can be done using the algorithm of \cite{Johnson2013}. The update in Step 3(c) has an explicit form in the case of $M=2$ groups  (see Section~\ref{sec:algdetails} of the SM). For $M \geq 3$ groups, we make use of the solution of  \cite{Hocking2011}.

   If the output of Algorithm~\ref{alg:simple} has the property that $\beta_{1j}=\ldots=\beta_{Mj}$ for some  $j$, $1 \leq j \leq p$, then we conclude that at the $j$th locus, the $M$ mean genomic phenotype profiles are identical. Due to numerical issues, however, we may not observe exact equality between $\beta_{mj}$ and $\beta_{m'j}$ for $m \neq m'$. Therefore, in practice, we set a threshold $\varepsilon$, and conclude that $\beta_{mj}$ and  $\beta_{m'j}$ are equal if the absolute difference between them is below $\varepsilon$. 
  The mean genomic phenotype profile for the $m$th group can be obtained from $\vs\theta_m$ in the output of Algorithm~\ref{alg:simple}.

In practice, it is computationally prohibitive to solve the JADE optimization problem \eqref{eq:jadeproblem_full} on genome-sized data. Therefore, we take a pragmatic approach: we segment the genome, and apply JADE to each segment in parallel.  In the methylation data application presented in Section 5, there is a natural segmentation that respects the biology of the problem. Other situations might require more arbitrary segmentation. Provided that the regions to be detected by JADE are short relative to the segmentation that we impose, we expect the segmentation to have little effect on the results.

\subsection{Tuning Parameter Selection}
\label{sec:cv}
The JADE optimization problem in (\ref{eq:jadeproblem_full}) involves two non-negative tuning parameters.   The parameter $\lambda$ controls the smoothness of the mean genomic phenotype profiles while $\gamma$ controls the amount of fusion between pairs of profiles. We take a two-stage approach to select $\lambda$ and $\gamma$ rather than performing a grid search over all combinations of values.

In both stages, cross-validation is performed by dividing the $pM$ data points $\bar{y}_{mj}$, $m\in \{ 1\dots M \}$, $j\in \{ 1\dots p \}$, into $l$ folds. For a given value of $m$, each fold contains a data point at every $l$th position, and the folds are   staggered so that all $m$ data points at a single position are not in the same fold. For example, if $M=2$, $p=10$, and $l=5$, then the first fold could contain $\bar{y}_{1,1}, \bar{y}_{1,6}, \bar{y}_{2,2},$ and $\bar{y}_{2,7}$.

 In the first stage,
  we set $\gamma=\infty$  in \eqref{eq:jadeproblem_full}; this amounts to combining all of the data into a single trend filtering problem. 
  We then perform cross-validation in order to select $\lambda$.
 
In the second stage of tuning parameter selection, we hold $\lambda$ fixed at the value selected in the first stage,  and select the tuning parameter $\gamma$ using cross-validation.
Additional details are provided in Section~\ref{sec:cv_gamma} of the SM. 
  
  In both stages of cross-validation, we apply the one-standard-error rule, selecting the largest tuning parameter value that has cross-validation error within one standard deviation of the minimum \citep{Hastie}. 
  
\section{Simulations}
\label{sec:sims}
In Section~\ref{sec:normal_sims}, we consider a setting in which the genomic phenotype is continuous-valued. In Section~\ref{sec:binom_sims}, we consider a setting  that is modeled after methylation sequence data.

\subsection{Normal Simulations}
\label{sec:normal_sims}

\subsubsection{Simulation Set-Up} \label{sec:simsetup}
\begin{figure}[!p]
\centering
\includegraphics[width=0.8\textwidth]{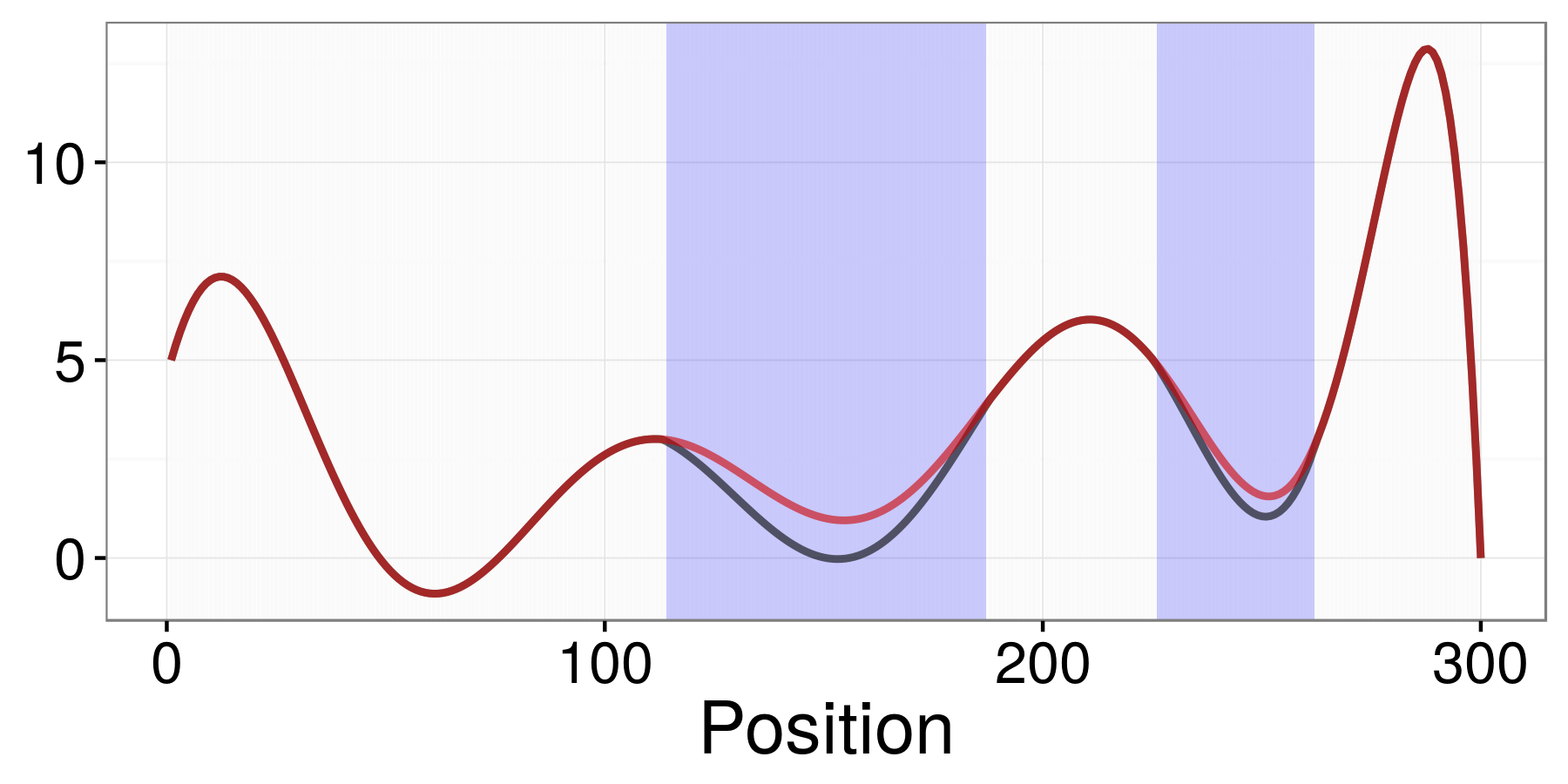}
\caption{Average group profiles for simulated data in Section~\ref{sec:sims}. The two profiles are separated in two regions highlighted in blue. In the white regions the two groups have the same mean. For binomial data simulations in Section~\ref{sec:binom_sims} these mean curves are scaled to range between 0 and 1.}
\label{fig:simmeans}
\end{figure}

 We simulate  $n=20$ observations, 10 in each of $M=2$ groups, at $p=300$ evenly spaced sites, $s_1,\ldots,s_p$. 
(We explore other values of the  sample size $n$ in Section~\ref{sec:ss_sims} of the SM.)
 The data for the $i$th observation in the $m$th group at the $j$th site is generated as
\begin{equation}
y_{imj} = f_{m}(s_{j}) + \epsilon_{imj},\label{eq:data-gen}
\end{equation}
where the functions $f_1$ and $f_2$ represent the mean genomic phenotype profiles for the two groups, and  are displayed in Figure~\ref{fig:simmeans}. The error terms $\epsilon_{imj}$  are generated in one of two ways: 
\begin{enumerate}
\item{\emph{Auto-regressive model.}} For $m=1,2$ and $i=1,\dots ,10$,
\begin{align*}
z_{imj}  \sim  & N(0, \sigma^{2}) \qquad  \mathrm{for \;\;} j = 1\dots 300,\\
\epsilon_{imj} = &
 \begin{cases}
z_{imj} \qquad & \mathrm{if \;\;} j = 1\\
z_{imj} + \rho z_{im(j-1)} \qquad & \mathrm{if \;\;} j > 1
\end{cases}.
\end{align*}
We consider values of $\sigma \in \lbrace 0.5, 1, 2\rbrace$ and $\rho \in \lbrace 0, 0.2, 0.4\rbrace$.
\item{\emph{Random effects model.}} For $m=1,2$, $i=1,\ldots,10$, and $j=1,\ldots,300$,
\begin{equation}
b_{im}  \sim  N(0, \sigma_{\mathrm{re}}^{2}), \qquad
z_{imj} \sim  N(0, \sigma^{2}), \qquad
\epsilon_{imj} =  b_{im} + z_{imj}.
\label{eq:RE}
\end{equation}
In this set-up, $b_{im}$  represents a mean shift for the $i$th observation in the $m$th group, such as one might expect as a result of a  batch effect. 
 We choose $\sigma$ and $\sigma_{\mathrm{re}}$  such that $\sigma^{2} + \sigma^{2}_{\mathrm{re}}=5$,   and the proportion of variance due to random effects, $\sigma^{2}_{\mathrm{re}}/\left(\sigma^{2} + \sigma^{2}_{\mathrm{re}}\right)$, takes on values of $0.05$, $0.1$, $0.15$, and $0.2$.
\end{enumerate}

\subsubsection{Methods for Comparison}\label{sec:normalmethods}
In this section, we compare JADE to three $t$-test based methods. These methods  decouple the tasks of estimating the mean genomic phenotype profiles for each of the $M$ groups, and testing for differences between the $M$ mean genomic phenotype profiles. These approaches assume that $M=2$.
\begin{enumerate}
\item A two-sample $t$-statistic is calculated at each site, without first smoothing the data. This approach is used by methylKit \citep{Akalin2012}. 
\item Each observation is smoothed using local likelihood, with the bandwidth chosen by generalized cross-validation. Then a two sample $t$-statistic is computed at each site, using the smoothed observations. BSmooth \citep{Hansen2012} uses this strategy with a fixed bandwidth optimized for methylation data. 
\item Each observation is smoothed using a quadratic smoothing spline, with the tuning parameter chosen by generalized cross-validation. Then a two sample $t$-statistic is computed at each site, using the smoothed observations.
\end{enumerate}
The third method is included in order to understand the impact of different smoothing strategies. 
  For all three methods, a threshold is chosen, and any site with a test statistic exceeding that threshold in absolute value is declared to have a different mean value between the $M$ groups.

We use our own implementation in \verb=R= of all three strategies, because methylKit and BSmooth are both implemented specifically for methylation count data, whereas the genomic phenotypes in this simulation study are continuous.
 We do not include the WaveQTL method of \cite{Shim2013} in our comparisons, as it requires the user to pass in pre-specified genomic regions, and does not provide a per-site assessment  of the association between genomic phenotype and category.
In our application of JADE, we set the weight matrices $A_{1}$ and $A_{2}$ in \eqref{eq:jadeproblem_full} to the identity. Tuning parameters were selected according to the procedure in Section~\ref{sec:cv}.

\subsubsection{Results}
\label{sec:normalresults}

We now compare the performances of JADE and the three $t$-test-based methods described in Section~\ref{sec:normalmethods}. 
 Before presenting these results, we briefly discuss  the calculation of false and true positives for each method.
 
 For a given value of $\gamma$ in the JADE optimization problem \eqref{eq:jadeproblem_full},  
  we declare a false positive if $\hat{\theta}_{1j} \neq \hat{\theta}_{2j}$ and $f_1(s_j) = f_{2}(s_j)$, and a true positive if $\hat{\theta}_{1j} \neq \hat{\theta}_{2j}$ and $f_1(s_j) \neq f_{2}(s_j)$. 
 We fit JADE at  around 100 values of $\gamma$, as described in Section~\ref{sec:cv_gamma} of the SM. For each value of $\gamma$ considered, we calculate a true positive rate and a false positive rate.

For a given $t$-statistic method and a given choice of threshold, we declare a false positive if the absolute value of the $t$-statistic for the $j$th site exceeds the threshold and $f_1(s_j) = f_{2}(s_j)$. We declare 
a true positive if the absolute value of the $t$-statistic for the $j$th site exceeds the threshold and $f_1(s_j) \neq f_{2}(s_j)$.
For each method, we calculate true positive and false positive rates  for a sequence of threshold values.

\begin{figure}[!h]
\centering
\includegraphics[width=\textwidth]{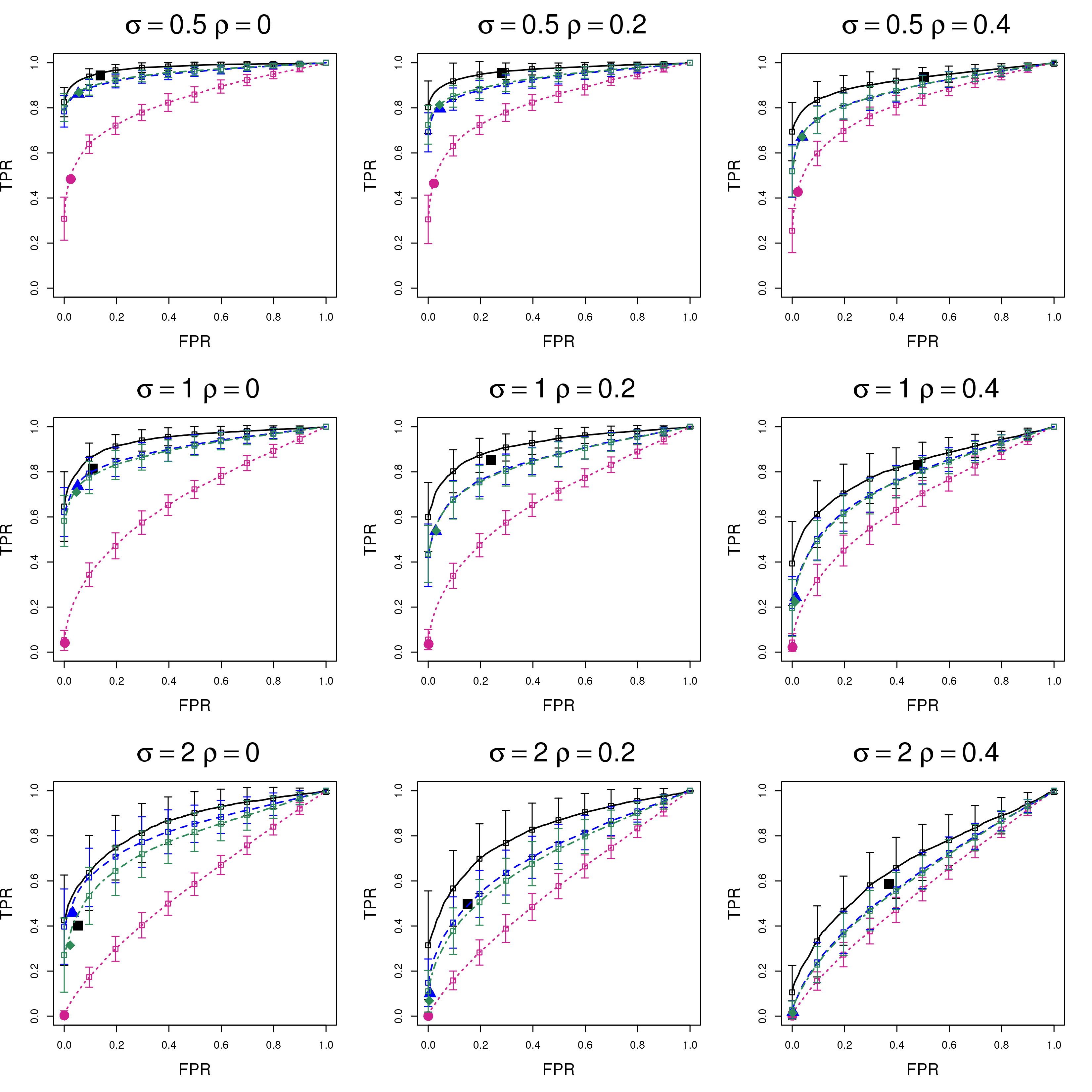}
\caption{Performance of JADE and competing methods in the normal auto-regressive model described in Section \ref{sec:simsetup}. Each panel displays results for a distinct value of $\sigma \in \{0.5,1,2\}$, and a value of $\rho \in \{0, 0.2, 0.4\}$.  
Lines show the average TPR for a fixed FPR, averaged over 100 simulations. The lengths of the vertical bars on either side of the curves equal one sample standard deviation of the TPR.
Points indicate average TPR and FPR achieved for JADE with the tuning parameter selected by cross-validation, and for the $t$-test approaches with an FDR threshold  of $10\%$. 
Methods shown are  JADE (\protect\includegraphics[height=.17cm]{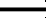},\protect\includegraphics[height=.17cm]{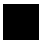} ), per-site $t$-tests applied to the raw data (\protect\includegraphics[height=.17cm]{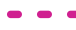}, \protect\includegraphics[height=.17cm]{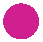}), and per-site $t$-tests after smoothing the raw data using splines (\protect\includegraphics[height=.17cm]{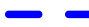}, \protect\includegraphics[height=.17cm]{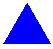}) and local likelihood (\protect\includegraphics[height=.17cm]{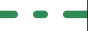}, \protect\includegraphics[height=.17cm]{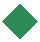}).
Results for the $t$-test with spline and local-likelihood smoothing are often nearly identical.
}
\label{fig:arsim}
\end{figure}

\begin{figure}[!h]
\centering
\includegraphics[width=\textwidth]{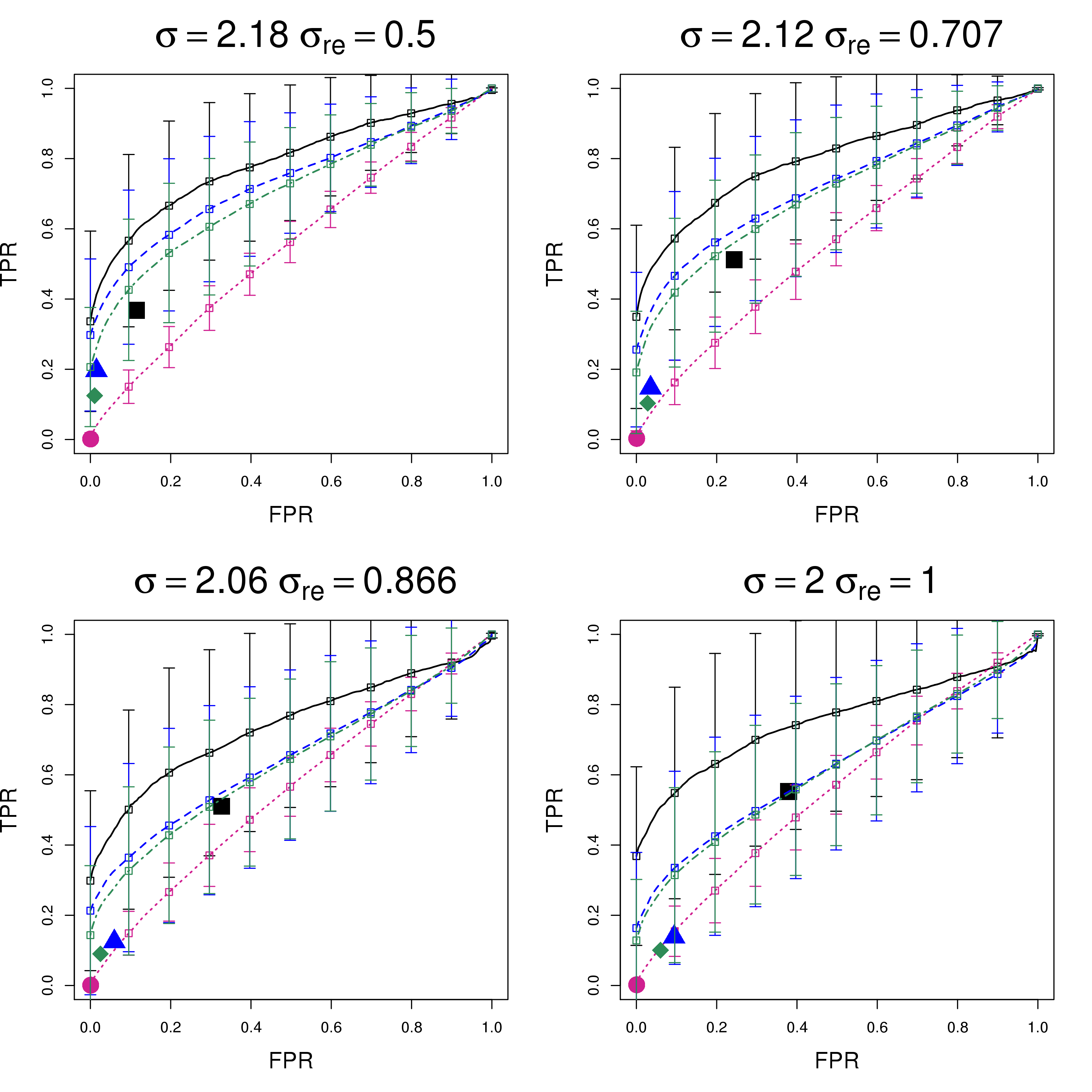}
\caption{Performance of JADE and competing methods in the normal random effects model described in Section \ref{sec:simsetup}. Each panel represents a different proportion of variation due to random effects. Additional details are as in Figure~\ref{fig:arsim}. }
\label{fig:resim}
\end{figure}

Figures~\ref{fig:arsim} and \ref{fig:resim} display the average true positive rate (TPR) as a function of the false positive rate (FPR) for JADE and the three $t$-test-based methods, for the two error structures described in Section~\ref{sec:simsetup}, averaged over {100} simulations. 
 Colored points indicate the average TPR and FPR achieved with tuning parameters selected via cross-validation for JADE, or using a false discovery rate (FDR)  of $10\%$ for the $t$-statistic-based methods, as calculated using SLIM  \citep{Wang2011}. 
 Details of the calculation of these curves are given in Section~\ref{sec:roc} of the SM.

In all settings,  JADE results in a higher TPR for any fixed FPR than the competing methods. 
 We expect JADE to perform well in the random effects setting because it pools observations within each group before smoothing, thereby averaging out individual-level random effects. The JADE framework does not, however, account for the dependence between errors seen in the auto-regressive simulations. These results show that JADE is robust, at least in this setting, to dependence between errors.
 
The $t$-statistic-based methods with an FDR cutoff of 10\% tend to be more conservative  than JADE with $\gamma$ chosen by cross-validation: that is, they yield fewer false positives and fewer true positives. The average FPR for JADE with $\gamma$ chosen by cross-validation increases for larger values of $\rho$ in the auto-regressive settings and $\sigma_{\mathrm{re}}$ in the random effects settings.

In Section~\ref{sec:rl_sims} of the SM, we evaluate JADE and the three $t$-test methods using a different approach, in which we treat contiguous blocks of associated sites as single discoveries.

\subsection{Binomial Simulations}
\label{sec:binom_sims}

\subsubsection{Simulation Set-Up and Methods for Comparison}\label{sec:binomsetup}
In this section, we use methylation sequence data to motivate our simulation set-up.   
DNA methylation is a chemical modification that can affect cytosine residues directly followed by guanine residues (CpGs).
In methylation sequencing experiments, DNA is fragmented, amplified, and bisulfite converted, a process in which non-methylated cytosines in CpGs are converted to uracil. These fragments are then sequenced, and the uracils and cytosines at each CpG are counted. Thus, at each CpG site we obtain two numbers: the number of sequenced fragments (reads) and the number of observed uracils (counts). We analyze methylation data in Section~\ref{sec:methylation}. In this section we consider a simple simulation mimicking the binomial character of methylation data.

 As in Section~\ref{sec:simsetup}, we simulate $n=20$ observations, 10 in each of $M=2$ groups at $p=300$  evenly spaced sites. 
 We generate the observed number of counts for the $i$th individual in the $m$th group at the $j$th site as  
\begin{align*}
c_{imj} \sim \text{Binom}(n_{imj}, p_{imj}).
\end{align*}
Section~\ref{sec:binom} of the SM describes the way in which $n_{imj}$, the total number of reads, is generated. No sites were permitted to have zero reads, as neither BSmooth \citep{Hansen2012} nor methylKit \citep{Akalin2012} can accommodate this.

 In order to generate the binomial probability $p_{imj}$, we first scaled and translated $f_1$ and $f_2$, the two mean genomic phenotype profiles displayed in Figure~\ref{fig:simmeans}, to take on values between $0$ and $1$. We then generated $p_{imj}$ according to a random effects model, as follows:

\begin{align*}
b_{im}  \sim  N(0, \sigma_{\mathrm{re}}^{2}), \hspace{10mm} p_{imj} =  \begin{cases} 0 \qquad & \text{if\;\;} f_{m}(s_{j}) + b_{im} < 0\\
1 \qquad  &\text{if\;\;}  f_{m}(s_{j}) + b_{im} > 1\\
f_{m}(s_{j}) + b_{im} & \qquad \text{otherwise}
\end{cases}.
\end{align*}
We consider values of $\sigma_{\mathrm{re}}\in \lbrace 0, 0.02, 0.05, 0.07\rbrace$.

We fit JADE using the observed proportions $y_{imj} = c_{imj}/n_{imj}$. 
Due to  the binomial mean-variance relationship as well as  the variable read depth, the variance of $y_{imj}$ is not constant across sites or observations. We
can estimate the variance of $y_{imj}$ as 
\begin{equation}
\hat{\sigma}_{imj}^{2} = \frac{y^{*}_{imj}(1-{y}^{*}_{imj})}{n_{imj}}, \hspace{10mm} \mathrm{where} \;\;
y^{*}_{imj} = \frac{c_{imj} + 0.5}{n_{imj} + 1}. 
\label{eq:binom_sd}
\end{equation}
Here, $y^{*}_{imj}$ differs from $y_{imj}$ in the inclusion of pseudo-counts to prevent estimates of zero variance. To accommodate these variance estimates in JADE, the  diagonal elements of the matrix $A_{m}$ in \eqref{eq:jadeproblem_full} were set to $1/\hat{\sigma}_{imj}$.

In what follows, we compare JADE to two existing methods 
  for analyzing methylation data, methylKit \citep{Akalin2012} and BSmooth \citep{Hansen2012}. These are methylation specific implementations of methods 1 and 2 in Section~\ref{sec:normalmethods}. 
  
  The local likelihood smoothing bandwidth is fixed in BSmooth, making its performance dependent on the spacing of the measurement sites. 
     For this comparison, we used a separation between sites of five units ($s_{1}=0, s_2 = 5, s_3=10,\dots$). This spacing is close to what might be expected of bisulfite sequencing data in which measurements are closely spaced but not made at every base-pair. %

\subsubsection{Results} 
\begin{figure}[!h]
\centering
\includegraphics[width=\textwidth]{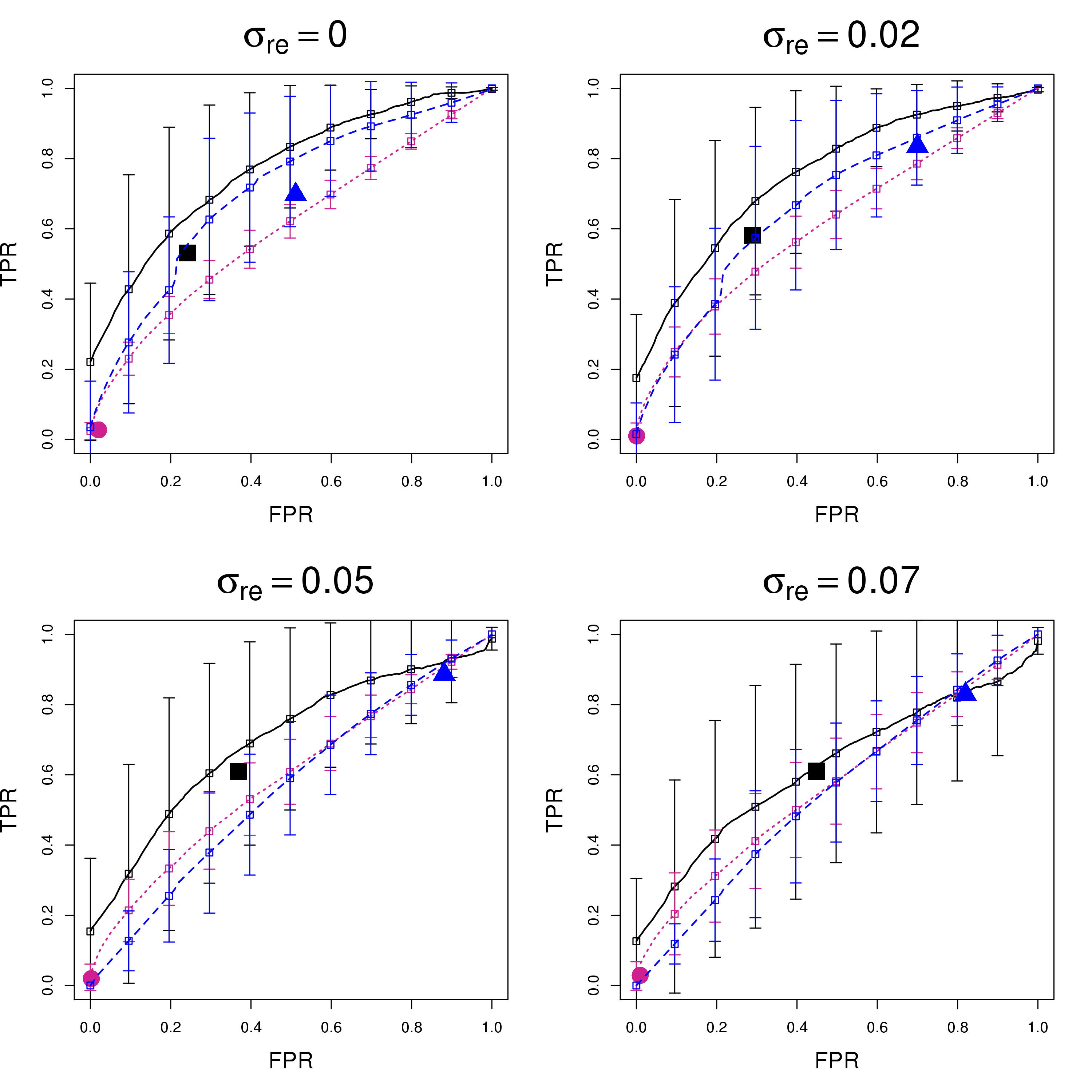}
\caption{Performance of JADE and competing methods in the binomial simulations described in Section \ref{sec:binomsetup}. Lines show average TPR for a fixed FPR over 100 simulations. The lengths of the vertical bars on either side of the curves equal one sample standard deviation of the TPR. Points indicate average TPR and FPR for JADE with the tuning parameter selected by cross-validation, and for the methylKit and BSmooth with an FDR threshold  of $10\%$. Methods shown are JADE (\protect\includegraphics[height=.17cm]{black.png}, \protect\includegraphics[height=.17cm]{blacksquare.png}), methylKit (\protect\includegraphics[height=.17cm]{pinkline.png}, \protect\includegraphics[height=.17cm]{pinkdot.png}), and BSmooth (\protect\includegraphics[height=.17cm]{blueline.png}, \protect\includegraphics[height=.17cm]{bluetriangle.png}).
 }
\label{fig:binom}
\end{figure}

Both BSmooth and methylKit produce  a score for each site, quantifying the evidence that the mean profiles differ at that location.  TPRs and FPRs for JADE, BSmooth, and methylKit were computed as described in  Section~\ref{sec:normalresults}. The results, averaged over {100} simulated data sets, are displayed in Figure~\ref{fig:binom}.
We find that JADE gives a higher TPR than BSmooth and methylKit for any fixed FPR in all settings. 

\section{Application to Methylation Data}

\label{sec:methylation}
 In this section, we apply JADE to DNA methylation patterns during three stages of skeletal muscle cell development (myoblast, myotube, and adult muscle cells), using reduced representation bisulfite sequencing data from the ENCODE project \citep{ENCODEProjectConsortium2012}. Methylation data were described at the beginning of Section~\ref{sec:binom_sims}.

These cell lines have been studied extensively: in particular,  ChIP-seq peaks, DNaseI peaks, and H3K27ac marks are also available. Therefore, we are able  to compare the set of differentially methylated regions (DMRs) detected by JADE with previous findings, and we can assess co-localization with other functional annotations in order to validate our results.  In what follows, we will make use of the fact that there is a developmental ordering to the three cell types in our data: myoblasts precede myotubes, which precede mature muscle cells.

\subsection{Analysis}
\label{sec:meth_analyis}

We compared DNA methylation in myoblasts, myotubes, and mature skeletal muscle cells. Three technical replicates from a single cell line are available for both myoblasts and myotubes, and two technical replicates are available  for mature skeletal muscle. 
We pooled technical replicates, and set the diagonal elements of the $A_m$ weight matrices in (\ref{eq:jadeproblem_full}) equal to the inverse of the standard deviation estimates in (\ref{eq:binom_sd}). In this analysis, we only examined chromosome 22.

 The locations at which DNA methylation can occur, CpG sites, are irregularly distributed throughout the genome. Since there is no biological reason to smooth across very long distances containing no CpG sites, this irregular spacing provides a natural way to segment the genome. 
   We divided the chromosome into segments such that neighboring CpG sites within a segment are separated by less than 2 kb, and the first and last CpG of each segment is measured in all three cell types. Segments with fewer than 20 CpGs were removed. This resulted in 477 segments with an average segment length of 3.0 kb and an average of 64 CpG sites per segment.
Running JADE on each of the 477 segments in parallel, with 5-fold cross-validation, required computing efforts equivalent to running 120 cores for two days.

Neither methylKit nor BSmooth can be directly applied to this data, since both methods are intended for a two-group comparison, and in this data set we have three groups.

\subsection{Results}
\label{sec:results}

\subsubsection{DMRs Identified by JADE}
We applied JADE to each of the 477 segments on chromosome 22, with $\lambda$ and $\gamma$ selected using 5-fold cross-validation as described in Section~\ref{sec:cv} and Section~\ref{sec:cv_gamma} of the SM, and with $\varepsilon=0.005$ as described in Section~\ref{sec:algo}.

We declared a DMR as any contiguous set of CpGs at which two or more profiles are separated in the JADE output. Adjacent DMRs separated by a single CpG are combined to form a single DMR.
We removed 48 regions containing 536 base-pairs (0.3\% of base-pairs in DMRs) in which two pairs of profiles are fused (separation $<\varepsilon$) while the third pair is un-fused (separation $>\varepsilon$), since such a pattern does not give a valid partition of the three profiles.

\begin{figure}[!p]
\centering
	\begin{subfigure}[b]{\textwidth}
	\caption{Raw Data}
		\includegraphics[width=\textwidth]{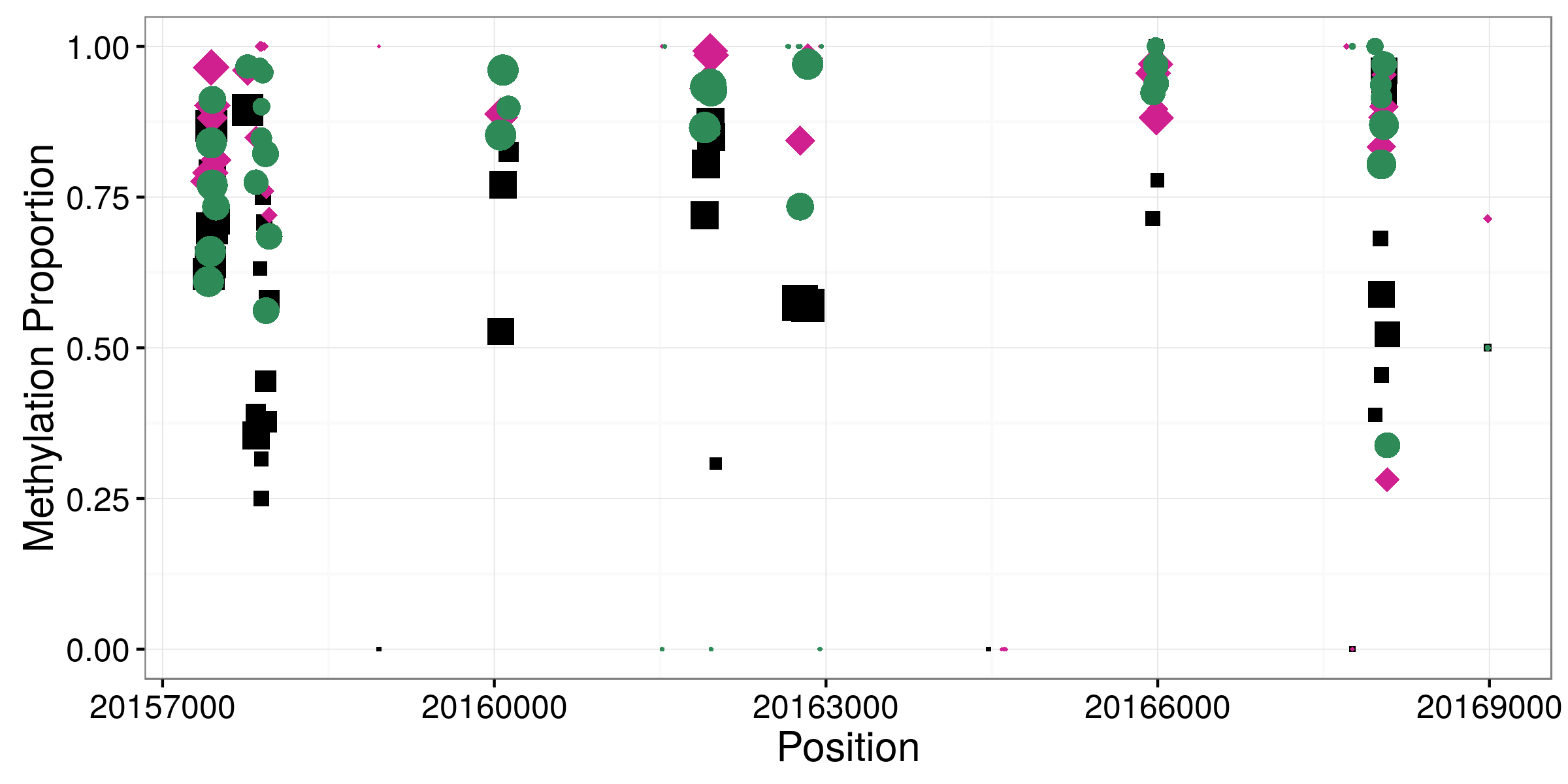}
		\label{raw52}
	\end{subfigure}\\
	\begin{subfigure}[b]{\textwidth}
	\caption{JADE Profile Estimates}
		\includegraphics[width=\textwidth]{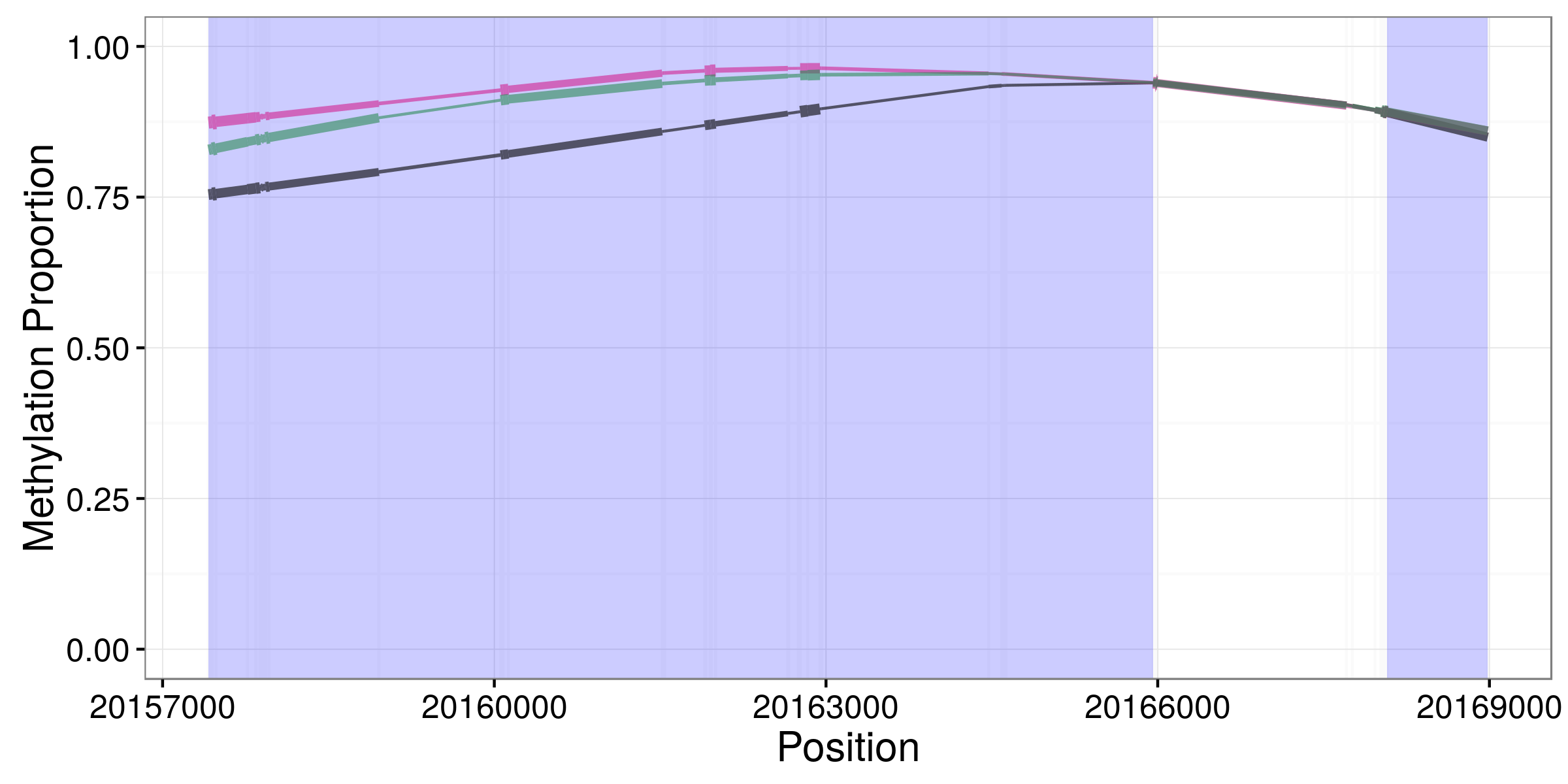}
		\label{jade52}
	\end{subfigure}
\caption{Results from one segment of the methylation data analysis, described in detail in Section~\ref{sec:results}.  Panel \ref{raw52} shows the raw data for myoblasts (\protect\includegraphics[height=.17cm]{greendiamond.png}), myotubes  (\protect\includegraphics[height=.17cm]{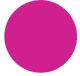}) and mature skeletal muscle (\protect\includegraphics[height=.17cm]{blacksquare.png}). Point size is proportional to the number of reads at each site.
Panel \ref{jade52} shows the three profiles estimated by JADE (top line: myotubes, middle: myoblasts, bottom: mature skeletal muscle). Blue shading indicates DMRs detected by JADE. Line width  is proportional to the number of reads in a 200 base-pair window.
}
\label{fig:w52}
\end{figure}

JADE identified 220 DMRs in 127 segments, {with an average length of 826 base-pairs}. 
   An example JADE fit is shown in Figure~\ref{jade52}. In this segment, two DMRs have been identified (shaded in blue).  
     In the DMR on the left, all three profiles are separated, while on the right, the myotube and myoblast profiles are fused. 
     
     A single DMR may contain multiple partitions of the profiles, though those in Figure~\ref{jade52} each contain only one partition.  The 220 DMRs identified by JADE can be divided into 380 sub-regions, each of which contains only a single partition of the profiles.

\subsubsection{Co-Localization of DMRs with Genetic and Epigenetic Landmarks}\label{sec:coloc}

 In order to assess the quality of the DMRs detected by JADE, we evaluate their overlap with  epigenetic annotations and genetic landmarks. 
  We expect the DMRs detected by JADE to be enriched for some of these genetic features.

  We consider epigenetic annotations obtained in myotubes and mature skeletal muscle, available from ENCODE. (Annotations obtained in myoblasts were not available.) These annotations include (i) transcription factor binding sites identified through ChIP-seq; (ii) active enhancer regions indicated by H3K27ac histone methylation; and (iii) DNase I hypersensitive sites, which mark open chromatin. 
 
We also consider three types of genetic landmarks: (i) CpG islands, annotated in the UCSC genome browser. Evidence suggests that methylation in these regions affects gene expression \citep{Bell2011,Illingworth2009}. (ii) CpG island shores, defined as the 2 kb flanking regions of islands. Using BSmooth, \citet{Irizarry2009} found that  DMRs between colon cancer and healthy colon cells tend to be located in CpG island shores.   (iii) The 2 kb flanking regions of transcription start sites (TSSs), annotated in the Gencode project \citep{Harrow2012} and available from ENCODE annotations. These 2 kb flanking regions serve as proxies for promoter regions, which are typically located immediately upstream of the TSS.

We tested whether the number of detected DMRs overlapping each genetic feature differed from what would be expected by chance. Details are given in Section~\ref{sec:meth_null} of the SM.
 We found that the DMRs identified by JADE are enriched in TSS flanking regions (Table~\ref{tab:results}).

Next, we restricted our analysis to the sub-regions of DMRs detected by JADE 
    that are consistent with increasing methylation over the course of development (Myoblast$\leq$Myotube$\leq$Mature)  or decreasing  methylation over the course of development (Myoblast$\geq$Myotube$\geq$Mature). These two groups account for approximately 58\% of all base-pairs in detected DMRs, as discussed in Section~\ref{sec:methylpattern} of the SM. 
  The results can be found in  Table~\ref{tab:gain_loss} of the SM. We found that the rate of overlap with CpG islands is higher in loss-of-methylation than in gain-of-methylation DMR sub-regions. This pattern is consistent with prior suggestions that demethylation plays a major role in up-regulating cell type specific gene expression over the course of development \citep{Segales2014,Hupkes2011}. We also found that the rates of overlap with both DNase-I hypersensitive sites and H3K27ac modifications are higher in gain-of-methylation than in loss-of-methylation DMR sub-regions.

\section{Discussion}
\label{sec:disc}

 In this manuscript we propose JADE, a flexible method for the analysis of genomic phenotypes measured in two or more conditions. JADE combines smoothing and group comparison into a single optimization problem, resulting in improved  power over competing methods.

 In addition to gains in power for simple comparisons of two groups, JADE offers a novel approach for analyzing data with respect to a categorical outcome. Although the BSmooth and methylKit frameworks could be extended to categorical outcomes by using a one-way ANOVA,
 categorical outcomes are typically analyzed by performing pairwise comparisons, as in \cite{Carrio2015}, or one-versus-all comparisons.  
 For example, tissue-specific DMRs have been identified by finding regions for which one cell type differs from the average over all other cell types \citep{Irizarry2009,ENCODEProjectConsortium2012}. 
 JADE is able to  identify DMRs across categorical outcomes, and determine the order and grouping of profiles within DMRs.

JADE is implemented as an \verb=R= package \verb=jadeTF=, currently available at the author's website \url{https://github.com/jean997/jadeTF}.

\section{Description of Supplementary Materials}
The reader is referred to the on-line Supplementary Materials for technical appendices.

\section{Funding}
D. W. was partially supported by a Sloan Research Fellowship, NIH Grant DP5OD009145, and NSF CAREER DMS-1252624. N.S. was supported by NIH Grant DP5OD019820. J.M. was supported by NIH Grant DP5OD019820 and NIH Grant F31HG008572. Methylation data analysis made use of resources provided by a Google Cloud Credits Award from Google Research.

\bibliographystyle{biorefs}
\bibliography{jade_bib}

\clearpage
\newpage

\clearpage
\newpage

\begin{table}
\caption{Overlap between detected DMRs and genetic features, for the methylation data analysis discussed in Section~\ref{sec:coloc}. 
`Total DMRs' is the number of DMRs that overlap each  genetic feature.
 `Fold' is the ratio of the observed number of overlapping DMRs to the number that would be expected by chance. `P-value' is the p-value based on a Fisher's exact test comparing the proportion of DMRs overlapping the genetic feature to the proportion expected to occur by chance (see Section~\ref{sec:meth_null} of the SM).
 }

\begin{tabular}{l|ccc}
Genetic Feature & Total (N=220) & Fold & P-Value \\
\hline
Cpg Islands & 119 (54.1\%) & 1.13 & 0.25 \\ 
  CpG Island Shores & 70 (31.8\%) & 0.99 & 1 \\ 
  Transcription Start Sites & 112 (50.9\%) & 1.32 & 0.011 \\ 
  TF Binding Sites & 25 (11.4\%) & 1.03 & 1 \\ 
  DNase I HS Sites & 95 (43.2\%) & 1.04 & 0.77 \\ 
  H3K27ac Modifications & 36 (16.4\%) & 0.77 & 0.22 \\
\end{tabular}

\label{tab:results}
\end{table}




\end{document}